\documentstyle{article}
\begin{document}
\textwidth=127mm
\textheight=210mm
\newcommand{\be}{\begin{equation}}
\newcommand{\ee}{\end{equation}}
\newcommand{\bea}{\begin{eqnarray}}
\newcommand{\eea}{\end{eqnarray}}
\newcommand{\beaa}{\begin{eqnarray*}}
\newcommand{\eeaa}{\end{eqnarray*}}
\newcommand{\qd}{\quad}
\newcommand{\qqd}{\qquad}
\newcommand{\npb}{\nopagebreak[1]}
\newcommand{\nn}{\nonumber}
\title{\bf Generic solutions for some
integrable lattice equations}
\author{L.V.Bogdanov \\ \em  Landau Institute for Theoretical Physics,
IINS,\\ \em
GSP-1 117940, 2 Kosygina, Moscow v-334, Russia\\
e-mail leonid@cpd.landau.free.net;leonid@itp.chg.free.net}
\date{}
\maketitle
\begin{abstract}
We derive the expressions for $\psi$-functions and
generic solutions of lattice principal chiral equations,
lattice KP hierarchy and hierarchy including lattice
N-wave type equations. $\tau$-function of $n$ free
fermions plays fundamental role in this context.
Miwa's coordinates in our case appear as the lattice
parameters.
\end{abstract}
\section{Introduction.}
In our talk we investigate the solutions of
lattice principal chiral field equations, lattice
KP hierarchy and other more general hierarchy,
including KP hierarchy and equations of N-wave type.
We believe that the equations that we study here
are known in one or another form and we direct the
reader to the literature on this subject
\cite{lattice}. Our approach uses the ideas developed
by Zakharov and Manakov \cite{Manakov} and also some
elements of Segal-Wilson Grassmanian technique \cite{SW}.
We omit in this text the part of our talk devoted to
solutions with special asymptotic behaviour and nonlocal
$\bar{\partial}$-problem because these results are partly
published and they are not connected with the lattice part
of the talk.

We show that for the lattice principal chiral field
equation the problem of
constructing the $\psi$-function reduces to the problem
of transposition of matrix rational functions, which corresponds
to the matrix Riemann problem in the continuous case.

For the continuous KP hierarchy the dynamics on the
Grassmanian is defined by the group $\Gamma_+$, consisting
of holomorphic maps
$g:D_0\rightarrow {\bf C}^{\times}$,
where $D_0$ is the unit disk. For the lattice hierarchy
the dynamics is defined by some rather simple subgroup of this group
and it is possible to derive the formulae for the $\psi$-function and
solutions
explicitely. The $\tau$-function for $n$ free fermions plays a
fundamental role in this context.
\section{Lattice principal chiral field equations.}
In frame of inverse scattering method the
principal chiral field equations are distinguished
as simple and fundamental object posessing very special properties.
Per instance, the Goursat type boundary problem for these equations
can be reduced to matrix Riemann problem for {\em arbitrary}
boundary conditions, and a kind of  nonlinear D'Alembert principle
takes place for these equations (see \cite{Krichever}). In this
paper we investigate the corresponding constructions for the
case of discrete and integrable versions of
principal chiral field equations.
We start from the discrete Zakharov-Shabat pair. The problem
of constructing the $\psi$-function for arbitrary
boundary conditions reduces in the discrete case to
the problem of
transposition of rational matrix functions and can be solved directly
through the system of linear equations.

Let us introduce first a lattice U-V pair
\bea
T_1 \psi(n,m,\lambda) =U(n,m,\lambda)\psi(n,m,\lambda),\nn\\
T_2 \psi(n,m,\lambda) =V(n,m,\lambda)\psi(n,m,\lambda),
\label{UV}
\eea
where $(T_1 f)(n,m)=f(n+1,m)$, $(T_2 f)(n,m)=f(n,m+1)$
The compatibility condition for the system (\ref{UV})
(the lattice Zakharov-Shabat system)
is obtained by
performing two shifts in the different order, it reads
\be
(T_2 U)V=(T_1 V)U.
\label{ZS}
\ee
Let U and V be rational matrix functions having
one simple pole
\bea
U=(1+{u \over \lambda -1}),\\
V=(1+{v\over\lambda+1}).
\eea
Then the condition (\ref{ZS}) gives the equations
\bea
T_1 v-v={1\over2}((T_1 v)u-(T_2u)v), \nn \\
T_2 u-u=-{1\over2}((T_1 v)u-(T_2u)v).
\label{equations}
\eea
We call system (\ref{equations}) the
lattice principal chiral field equations.

Let us take a Goursat problem for the system of equations
(\ref{equations})
\bea
u(0,m)=u_0 (m),\\
v(n,0)=v_0 (n),
\eea
and consider a problem of constructing the function $\psi$
(and, consequently $u$ and $v$) for all $n,m\geq 0$.
We normalize $\psi$ by the condition
$$
\psi (0,0)=1.
$$
We can reach the point $(n,m)$ by different routes. Let us treat two
ways by the sides of the rectangular $(0,0),(0,M),(N,0),(N,M)$.
These two ways should give identical result for the function
$\psi$
\bea
&&\psi (N,M,\lambda)=\nn \\
&&\prod_{i=1}^N (1+{u(i,M)\over \lambda -1})
\prod_{i=1}^M (1+{v(0,i)\over \lambda +1})=
\prod_{i=1}^M (1+{v(N,i)\over \lambda +1})
\prod_{i=1}^N (1+{u(i,0)\over \lambda -1})
\label{ways}
\eea
Equality (\ref{ways}) may be considered as an equation for the
rational matrix functions $X(\lambda)$, $Y(\lambda)$
\bea
&&X(\lambda)
=\prod_{i=1}^N (1+{u(i,M)\over \lambda -1})=1+\sum_{i=1}^N
{x_i\over (\lambda -1)^i}\nn\\
&&Y(\lambda)
=\prod_{i=1}^M (1+{v(N,i)\over \lambda +1})
=1+\sum_{i=1}^M
{y_i\over (\lambda +1)^i},
\nn
\eea
it can be written in a form
\be
A(\lambda)X(\lambda)=B(\lambda)Y(\lambda),
\label{ways1}
\ee
where the functions $A(\lambda)$, $B(\lambda)$ are defined
by the boundary conditions; these functions are given by the
expressions
$$
A(\lambda)=\prod_{i=1}^M (1+{v(0,i)\over \lambda +1}),\quad
B(\lambda)=\prod_{i=1}^N (1+{u(i,0)\over \lambda -1}).
$$
Then, taking residues of different orders of equation (\ref{ways1}),
we obtain the system of $(N+M)$ linear equations for $(N+M)$ matrices
$x_i , y_i$. Solving this system of equations, we find the expression
for the function $\psi(N,M,\lambda)$ through the boundary conditions.

So the problem (\ref{ways1}) is used in the lattice case to construct
the $\psi$-function instead of the Riemann problem in the continuous
case. It reduces the Goursat problem for the lattice principal chiral
field equations (\ref{equations}) to the system of linear equations.
We will see later that reduction of integral equations used in
the continuous case to systems of linear equations in the lattice
case is quite a common feature.
\section{Lattice KP hierarchy}
We construct lattice KP hierarchy generalizing the technique that
was developed by Zakharov and Manakov for (2+1)-dimentional
dressing method and using some elements of Segal-Wilson
Grassmanian technique. The Grassmanian $G(H)$, where $H$ is the
Hilbert space $L^2(S^1)$ with the coordinate $\lambda$ on the
unit circle, is defined as a space of subsets $W$ that are
(in some precise sence) comparable to $H_+$ (the subspace
with the basis $1,\lambda,\lambda^2,...$) \cite{SW}.

There is a correspondence between KP hierarchy and the evolution
on the Grassmanian \cite{SW}. The evolution on the Grassmanian is
introduced in the following way
\be
g^{-1}W(g)=W_0,
\label{evolution}
\ee
where $g$ is the group element acting on the Grassmanian,
$g\in \Gamma_+$ (see the introduction),
$W(g)$ is the point of Grassmanian depending on the group element
and $W_0$ is the initial point parametrizing the solution, we
suppose it to be transversal to $H_-$, where $H_-$ is a space
with the basis $\lambda^{-i}$, $i>0$.

For the KP hierarchy the group element is taken in the form
\be
g^{-1}=\exp\sum_{i=1}^{\infty}\lambda^i t_i .
\label{exponent}
\ee
Let some function $\psi(t_i,\lambda)=\psi({\bf t},\lambda)$ belong to
$W(g)=W({\bf t})$.  Then
\be D_i\psi=({\partial\over\partial t_i}
+\lambda^i)\psi\in W(g),
\label{Manakov}
\ee
$$
u({\bf t})\psi\in W(g)
$$
(it easily follows from (\ref{evolution}) and
(\ref{exponent}), indeed, $g^{-1}\psi\in W_0\Rightarrow \partial_i
(g^{-1}\psi)
\in W_0\Rightarrow D_i\psi\in W(g)$).  So the operators $D_i$, $u$ form
the set of generators of the Manakov ring (we use the term Manakov ring
for the ring of operators possessing the property $DW(g)\in W(g)$).  Using
this ring, we can derive the set of linear operators defining the KP
hierarchy for $W$ transversal to $H_-$
\be (D_i-D_1^i)\psi(\lambda ,{\bf t})=
\sum_{k=0}^{i-2}u_k ({\bf t})D_1^k\psi,
\label{operators}
\ee where
$\psi\in W$, $\psi=1+\sum_{i=1}^{\infty} v_i({\bf t})\lambda^{-i}$ ,
potentials $u_i$ in the expressions (\ref{operators}) are defined through
the coefficients $v_i$ by taking the condition that the coefficients of
expansion of (\ref{operators}) in $\lambda$ are equal to zero for
$\lambda^n$, $n\geq 0$. Then due to transversality of $W$ expressions
(\ref{operators}) are equal to zero. One can easily go in
(\ref{operators}) from $D_i$ to partial derivatives $\partial_i$ by the
transform \be \psi\rightarrow g\tilde \psi \label{transform} \ee

For the lattice KP hierarchy the point of Grassmanian $W$ is a
function of integer parameters $W(n_1, n_2,...)=W({\bf n})$,
$\sum_{i=1}^{\infty}|n_i|<\infty$, the
dependence on this parameters being defined by the group element
\be
g=\prod_{i=1}^{\infty}(1-l_i \lambda^i)^{n_i},
\label{group1}
\ee
where $l_i$ is the step of the lattice in the corresponding
direction.
Now Manakov ring is generated by the operators
$$
D_i=(1-l_i \lambda^i)T_i,
$$
$$
D_iW\in W.
$$
Below we will use equivalent operators
$$
D_i=\Delta_i -\lambda^i T_i,
$$
$$
(\Delta_i f)(n_i)={f(n_i+1)-f(n_i)\over l_i},
$$
the continuous limit in this case is more transparent.
And we derive the analogue of expressions (\ref{operators})
\be
(D_i-D_1^i)\psi(\lambda ,{\bf t})=
\sum_{k=0}^{i-1}u_k ({\bf t})D_1^k\psi,
\label{operators1}
\ee
We can go from operators $D_i$ to operators $\Delta_i$
by the transform
\be
\psi\rightarrow g\tilde \psi.
\label{transform1}
\ee
To derive equations of lattice KP hierarchy as compatibility
conditions for equations (\ref{operators1}), one should use
a `Leibnitz rule' for difference operators $\Delta$:
\be
\Delta_i(u\psi)=(\Delta_i u)\psi+(T_i u)\Delta_i\psi.
\ee
To construct the solution of lattice KP hierarchy, we should
define the function $\psi({\bf n},\lambda)$ using $W_0$
as initial data. To do this,
we will use the formula for the action of the group element
\be
g=\prod_{i=1}^N(1-{\lambda\over z_i})
\label{group}
\ee
on the $\tau$-function, which can be derived for example in frame of
free fermion model of the Grassmanian \cite{Miwa}. Let
$f_i(\lambda)=\lambda^i+\sum_{k=1}^{\infty}v_{ik}\lambda^{-k}$ ,
$i\geq 0$ be the basis of $W_0$. Then this  formula reads
\be
\tau (z_1,...,z_n)={\det(f_i(z_j))\over \det((z_j)^i)},
\label{tau}
\ee
where $z_i$ are supposed to lie on the unit circle.
The parameters $z_i$ are often called Miwa's coordinates.
Let us suppose that the
functions $f_i(\lambda)$ are holomorphyc outside the
unit disk. Then the function (\ref{tau}) is defined
for arbitrary $z_i$ outside the unit disk.
The function $\tau(z_1,...,z_N)$ is symmetrical with
respect to its variables.
The $\psi$-function can be expressed through the
$\tau$-function by the formula
$$
\psi(g,\xi)=\tau(g\times (1-{\lambda\over\xi}))(\tau(g))^{-1}
$$
So the $\psi$-function for the group element (\ref{group}) is
\be
\psi(g,\xi)={\tau(z_1,...,z_n, z_{n+1}=\xi)\over \tau(z_1,...,z_n)},
\label{psi}
\ee
where we use the expression (\ref{tau}).

To find the $\psi$-function for the group element
(\ref{group1}),
we should use the function $\tau(z_1,...,z_n)$ with coinciding
parameters $z_i$.
Though the denominator of the function (\ref{tau}) has zero in this case,
it is compensated by the zero of enumerator and the function
has the limit when the parameters coincide
(we do not give the proof of this statement here).
It is not difficult to
find the expression for this limit in every particular case,
it will include the derivatives of basic functions $f_i$ with respect
to $\lambda$. In general case it is only a question of notations.
So we suggest the function $\tau(z_1,...,z_N)$ to be
defined for arbitrary
parameters $z_n$ outside the unit disk.

The formula (\ref{psi}) expresses the $\psi$-function at some point
through the `initial data' -- the basic functions at the initial
point and gives the solutions of lattice KP-hierarchy for
positive shifts ($n_i>0$). For example, the $\psi$-function
at the point $n_1=3, n_2=1, n_3=1, n_i=0\: i>3$ for the hierarchy
(\ref{operators1})
is given by the
formula
$$
\psi(\lambda)={\tau({1\over l_1},{1\over l_1},
{1\over l_1},{1\over\sqrt{l_2}},-{1\over\sqrt{l_2}},
{1\over (\root{3}\of{l_3})_1} ,
{1\over (\root{3}\of{l_3})_2},{1\over (\root{3}\of{l_3})_3},
\lambda)\over \tau({1\over l_1},{1\over l_1},
{1\over l_1},{1\over\sqrt{l_2}},-{1\over\sqrt{l_2}},
{1\over (\root{3}\of{l_3})_1} ,
{1\over (\root{3}\of{l_3})_2},{1\over (\root{3}\of{l_3})_3})} ,
$$
the $\tau$-function is given by the expression (\ref{tau}).
To find the solution at the arbitrary point,
we need some modification of the technique which will be introduced
in the next section.
\section{Solutions for the generic lattice equations}
The formulae for the solutions of the
hierarchy that we introduce in this part
are based on the action of the group element
\be
g^{-1}=\prod_{i=1}^N({\lambda -z_i\over \lambda- \hat{z}_i}).
\label{group3}
\ee
To calculate the action of this group element explicitly
we will introduce a kind of special basis in the
Grassmanian, making some assumptions about analytical
properties of vector space $W$.

Let
$$f_i=\lambda^{-i-1}+\sum_{k=0}^{\infty}v_{ik}\lambda^k,\qquad
i\geq 0$$
be the basis of $W$ (here for technical reasons
we work `in the neighborhood of zero', so now zero
corresponds to infinity of the previous sections,
and $\psi$-function of the previous section transforms
now to ${1\over\lambda}\psi({1\over\lambda})$). Zeroes
and poles of the group element (\ref{group3}) are
suppozed to lie in the unit disk.
Let us suppose that the series
$$\phi(\lambda,\mu)=\sum_{k=0}^{\infty}
\sum_{i=0}^{\infty}v_{ik}\lambda^k \mu^i$$
defines
a holomorphyc functions of two variables inside the
unit circle. Then the function
\be
\psi(\lambda,\mu)={1\over \lambda-\mu}+\phi(\lambda,\mu)
\label{function}
\ee
belongs to $W$ as a function of $\lambda$ for
$\lambda,\mu$ inside the unit circle. So there is
a correspondance between the point of Grassmanian
$W$ and the function of two complex variables
$\psi(\lambda,\mu)$. The function $\psi(\lambda,\mu)$
plays a role of the basic function which defines the
Grassmanian point $W$. To find the usual basis of $W$,
one should take derivatives of $\psi(\lambda,\mu)$
with respect to $\mu$ at the point $\mu=0$. Then
\be
f_i(\lambda)=({\partial\over\partial\mu})^{i-1}
\psi(\lambda,\mu)|_{\mu=0}.
\label{basis}
\ee

It is easy to check that in this case $W$ contains a
meromorphyc function (not unique in a general case)
for arbitrary given divisor of poles inside the unit
circle. We also suggest that $W$ is transversal to the space
of functions analytic inside the unit circle. Using these
properties and existence of the basic function
$\psi(\lambda,\mu)$ generating $W$, it is possible to
calculate the action of the group element
(\ref{group3}) explicitly in terms of the
basic function $\psi(\lambda,\mu)$.

Let the basic function (\ref{function}) be defined at the initial point.
The action of the group element (\ref{group3}) on the Grassmanian
is defined by the property (\ref{evolution}). Using this definition,
we search for the basic function (\ref{function}) in the form
$$\psi(\lambda,\mu)=g\psi_0(\lambda,\mu)-
\sum_{i} f_i(\mu)g\psi_0(\lambda,\mu_i).
$$
The simple trick is to use zeros of the function $g(\lambda)$
as the points $\mu_i$. Then, after simple calculations,
we derive the expression for the basic function $\psi(\lambda,\mu)$
in the space $W(g)$ in terms of initial basic function
\bea
&&\psi(g,\lambda,\mu)=\psi(z_1,...,z_N;\hat z_1,...,\hat z_N;
\lambda,\mu)=\label{function1}\\
&&\Phi(\lambda,\mu)-\sum_{i=1}^N\sum_{j=1}^N
(\Phi(\lambda,\mu)(\mu-\hat z_i))|_{(\lambda ,\hat z_i)}((\Phi)^{-1})_{ij}
(\Phi(\lambda,\mu)(\lambda-z_j))|_{(z_j ,\mu)},\nn
\eea
where
$$
\Phi(\lambda,\mu)=(\prod_{i=1}^N{(\lambda-\hat z_i)(\mu-z_i)
\over(\lambda-z_i)(\mu-\hat z_i)})\times \psi_0(\lambda,\mu),
$$
and the matrix $(\Phi)$ is
$$
(\Phi)_{ij}=(\Phi(\lambda,\mu)(\mu-\hat z_i)(\lambda-z_j))|_
{\lambda=z_j,\mu=\hat z_i}.
$$
This procedure works also in the case when $g(\lambda)$ is arbitrary
rational function with zeroes and poles only inside the unit disk.
So in usual terms it defines the action of
arbitrary rational function $g\in \Gamma_+$ on the Grassmanian
(zero in this section corresponds to infinity in usual
notations).

The function $\psi$ (\ref{function1}) is symmetrical with respect
to variables $z_i$ and also with respect to variables $\hat z_i$.
When some poles coincide, one should take corresponding limit of
this expression. We suggest that the function $\psi$
(\ref{function1}) is defined for arbitrary $z_i$,
$\hat z_i$ inside the unit circle (though we don't
prove this statement here).
The formula (\ref{psi}) from the previous section
may be obtained from the expression (\ref{function1}) in the
limit $\hat z_i=0,1\leq i\leq N$.

To introduce the corresponding hierarchy of integrable
lattice equations, we consider the group element
\be
g^{-1}=\prod_{i=1}^{\infty}(1+l_i K_i(\lambda))^{n_i},
\ee
where $K_i$ are arbitrary rational function on the
complex plane decreasing at infinity,
$l_i$ are the steps of the lattice,
$\sum_{i=1}^{\infty}|n_i|<\infty$. This group
element may be considered as degeneration of the group element
(\ref{group3}). The generators of Manakov ring
in this case are
\be
D_i=\Delta_i +K_iT_i.
\ee
The generic equations in this hierarchy
correspond to the operators $D_i$ with simple and
distinct poles. We use the functions $K_{i}(\lambda)$ :
\begin{equation}
K_{i}(\lambda)=\sum_{\alpha=1}^{n_{i}}\frac{a_{i}^{\alpha}}
{\lambda - \lambda_{i}^{\alpha}},
\end{equation}
where $a_{i}^{\alpha},\lambda_{i}^{\alpha} \in {\bf C}$,
$1 \leq \alpha \leq n_{i}$, $\lambda_{i}^{\alpha} \neq
\lambda_{j}^{\beta}$.
Then in the points of transversality
of the grassmanian the function $\psi(\lambda,\mu)$ satisfy the relations
\begin{eqnarray}
D_{i} \psi(\lambda,\lambda_j^{\beta},{\bf n}) -
K_{i}(\lambda_{j}^{\beta})\psi(\lambda,\lambda_j^{\beta},{\bf n})-
(T_i\psi(\lambda_j^{\beta},\lambda_i^{\alpha},{\bf n}))
\psi(\lambda,\lambda_
i^{\alpha}, {\bf n})=0,
\label{N}
\end{eqnarray}
where $i \neq j$ , summation over $\alpha$ is understood.
The leading order of expansion of the relation (\ref{N}) as
$\lambda \rightarrow \lambda_{k}^{\gamma}$, $i\ne j \ne k$
yields the equation
\begin{eqnarray}
\Delta_i \psi_{jk}^{\beta \gamma}+
K_{i}(\lambda_{k}^{\gamma})T_i\psi_{jk}^{\beta \gamma}
-K_{i}(\lambda_{j}^{\beta})\psi_{jk}^{\beta \gamma}-
(T_i\psi_{ji}^{\beta \alpha})a_{i}^{\alpha} \psi_{ik}^{\alpha
\gamma}=0,\label{NW}\\
\psi_{jk}^{\beta \gamma}=\psi(\lambda_k^{\gamma},
\lambda_j^{\beta},{\bf n})\nn
\end{eqnarray}
summation over $\alpha$ is understood.
If the different permutations $ijk$ and substitutions of
the indices $\beta ,\gamma$ are taken into account, (\ref{NW})
is a closed set of equations for the functions
$\psi_{ji}^{\beta \alpha}({\bf n})$ for chosen $ijk$.

Expression (\ref{function}) gives the solutions
for the system of equations (\ref{NW}) and also for the
lattice KP hierarchy at the arbitrary point. The function
$\psi_0(\lambda,\mu)$ defines the point of the Grassmanian
and plays the role of spectral data.

Let us restrict ourselves to the case of one pole
for the sake of simplicity.
The functions $K_i(\lambda)$ in this case are
$$
K_i(\lambda)={1\over \lambda-\lambda_i},
$$
the group
element, corresponding to the shift $T_1^m T_2^n T_3^p$
is
$$
g^{-1}=\left({\lambda-(\lambda_1-l_1)\over\lambda-\lambda_1}\right)^m
\left({\lambda-(\lambda_2-l_2)\over\lambda-\lambda_2}\right)^n
\left({\lambda-(\lambda_3-l_3)\over\lambda-\lambda_3}\right)^p
$$

So this group element is a special case of the group element
(\ref{group3}) with coinciding poles. To find the corresponding
function $\psi(\lambda,\mu)$, we should use the formula
(\ref{function}) for $N=m+n+p$, where $m$ variables
$z_i$ are equal to $\lambda_1$, $n$ variables to $\lambda_2$,
$p$ variables to $\lambda_3$, and $m$ variables $\hat z_i$
to $\lambda_1-l_1$, $n$ variables to $\lambda_2-l_2$,
$p$ variables to $\lambda_3-l_3$.

{\bf Remarks.} The technique developed here works also in the
case of q-difference equations. The author is planning to publish
corresponding results in the nearest future.

{\bf Acknowledgements}

The author is grateful to the organizers of NEEDS-93 for
financial support and for creating a very stimulating atmosphere
at the conference.

\end{document}